\documentclass{article}

\usepackage[preprint, nonatbib]{neurips_2023}

\usepackage[utf8]{inputenc} 
\usepackage[T1]{fontenc}    
\usepackage{hyperref}       
\usepackage{url}            
\usepackage{booktabs}       
\usepackage{amsfonts}       
\usepackage{nicefrac}       
\usepackage{microtype}      
\usepackage{xcolor}         

\usepackage{tikz}
\usetikzlibrary{tikzmark,decorations.pathreplacing,calligraphy}

\usepackage{amsmath,amsthm,amssymb,amsfonts}
\usepackage{bm}
\usepackage{multirow}
\usepackage{algorithm}
\usepackage{algpseudocode}

\usepackage{arydshln}

\newcommand{\norm}[1]{\left\lVert#1\right\rVert}

\title{Towards Generalizable Data Protection With Transferable Unlearnable Examples}

\begin{document}

\author{Bin Fang$^{1*}$\quad  Bo Li$^{2*}$\quad Shuang Wu$^{2}$\quad  \\ \textbf{Tianyi Zheng}$^{1}$\quad 
 \textbf{Shouhong Ding}$^{2}$\quad \textbf{Ran Yi}$^{1}$\quad  
\textbf{Lizhuang Ma}$^{1}$
\vspace{0.2cm}\\ 
\vspace{0.2cm}
$^1$Shanghai Jiao Tong University \qquad $^2$Youtu Lab Tencent \\
}

\maketitle
\renewcommand{\thefootnote}{\fnsymbol{footnote}}
\footnotetext[1]{Both authors contributed equally to this work. Work done during Bin Fang's internship at Tencent Youtu Lab and Bo Li is the project lead.}

\begin{abstract}

Artificial Intelligence (AI) is making a profound impact in almost every domain. One of the crucial factors contributing to this success has been the access to an abundance of high-quality data for constructing machine learning models. Lately, as the role of data in artificial intelligence has been significantly magnified, concerns have arisen regarding the secure utilization of data, particularly in the context of unauthorized data usage. To mitigate data exploitation,  data unlearning have been introduced to render data unexploitable. However, current unlearnable examples lack the generalization required for wide applicability. In this paper, we present a novel, generalizable data protection method by generating transferable unlearnable examples. To the best of our knowledge, this is the first solution that examines data privacy from the perspective of data distribution. Through extensive experimentation, we substantiate the enhanced generalizable protection capabilities of our proposed method.
\end{abstract}

\section{Introduction}

Over the last decade, the field of Artificial Intelligence (AI) has experienced significant advancements, leading to a substantial impact on nearly every domain. One of the crucial factors contributing to this success has been the access to an abundance of high-quality data for constructing machine learning models. Lately, the importance of data in AI has been greatly emphasized.
Many major AI breakthroughs in natural language processing ~\cite{DBLP:conf/naacl/DevlinCLT19,DBLP:journals/corr/abs-2303-08774}, computer vision~\cite{DBLP:conf/cvpr/RombachBLEO22,DBLP:conf/mindtrek/Oppenlaender22a} and computational biology~\cite{jumper2021highly} have been realized only after obtaining the appropriate training data. As the role of data in artificial intelligence has been significantly magnified, concerns have arisen regarding the secure utilization of data. On one hand, some technology companies utilize unauthorized private data to train their commercial models~\cite{openai,midjounary&sd,midjounary1}; on the other hand, there are enterprises that wish to protect their data assets, ensuring that they are not exploited by competitors for model training, thereby maintaining a leading advantage in their own models~\cite{DBLP:journals/corr/abs-2103-02683}.

In order to prevent unauthorized use of data, unlearnable examples have emerged as a promising countermeasure~\cite{Availability-Attack}. Recent research indicates that by injecting imperceptible noise into data, the performance of models employing such poisoned data can be significantly impaired~\cite{AA-1, TAP2021, NTGA2021, EM2021, LSP2022, CUDA2023, UC2023}. 
However, the majority of methods, including LSP~\cite{LSP2022}, CUDA~\cite{CUDA2023}, TAP~\cite{TAP2021}, NTGA~\cite{NTGA2021}, and EM~\cite{EM2021}, are susceptible to adversarial training. Specifically, unlearnable examples generated by LSP~\cite{LSP2022} and CUDA~\cite{CUDA2023} are strictly related to pixel values, rendering them non-robust against adversarial training. Furthermore, surrogate models of TAP~\cite{TAP2021}, NTGA~\cite{NTGA2021}, and EM~\cite{EM2021} undergo standard training, which implies that these methods are also vulnerable to adversarial training. Consequently, REM~\cite{REM2022} is introduced to generate robust unlearnable examples. However, REM exhibits limited concern for the generalizability of unlearnable examples.

In this paper, we propose a generalization-capable methodology for generating robust and transferable unlearnable examples by taking data distribution into account. Acknowledging that deep neural networks strive to discern the underlying data distribution, we aim to induce data collapse, thereby hindering models from effectively extracting information. Moreover, as data collapse is inherently associated with the data itself, embracing this perspective enables us to augment the generalization capabilities of model-based methods.

\begin{figure}[htbp]
\label{fig:effects}
\begin{minipage}[t]{0.45\linewidth}
\centering
\includegraphics[scale=0.4]{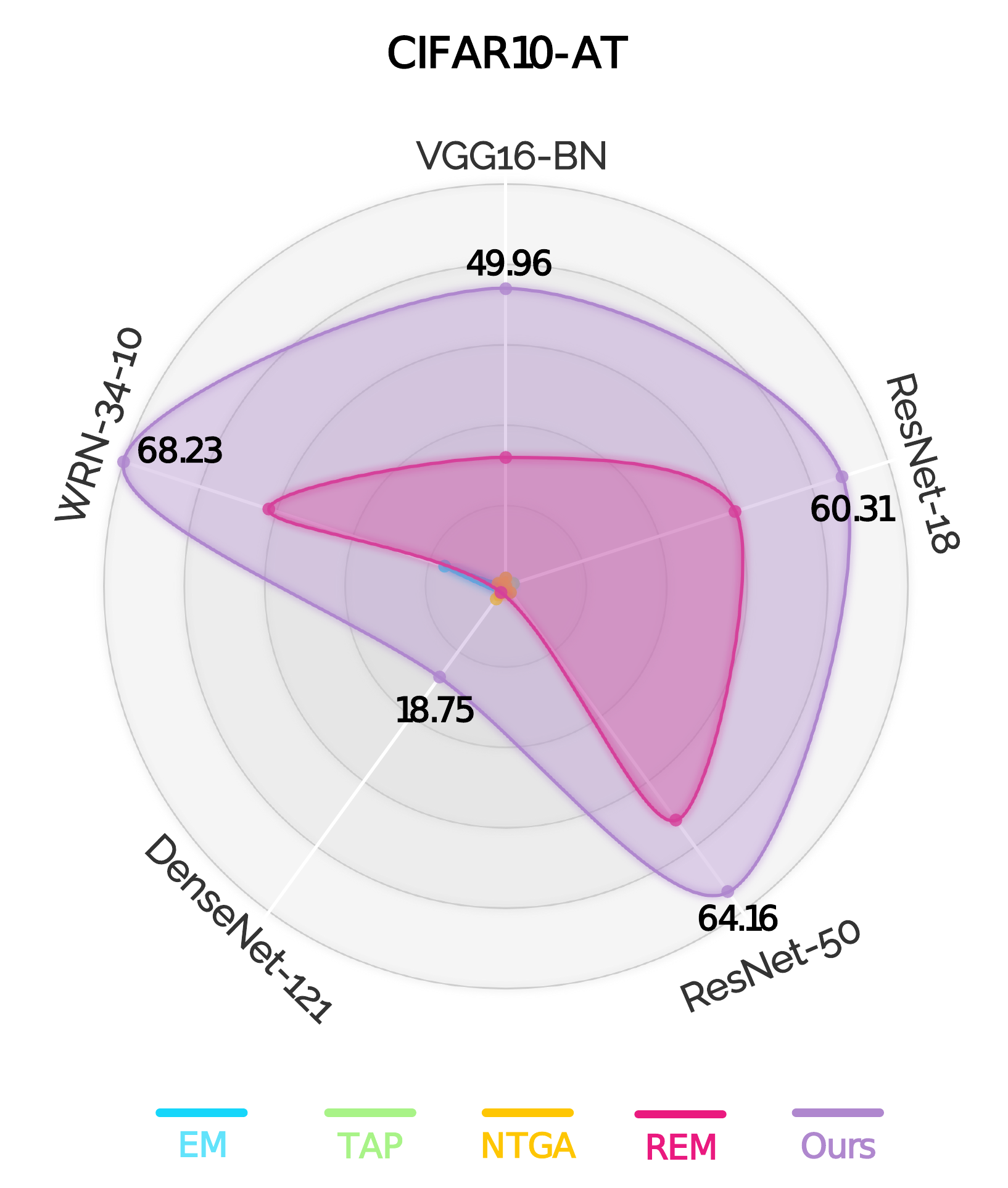}

\end{minipage}%
\hspace{10mm}
\begin{minipage}[t]{0.45\linewidth}
\centering
\includegraphics[scale=0.4]{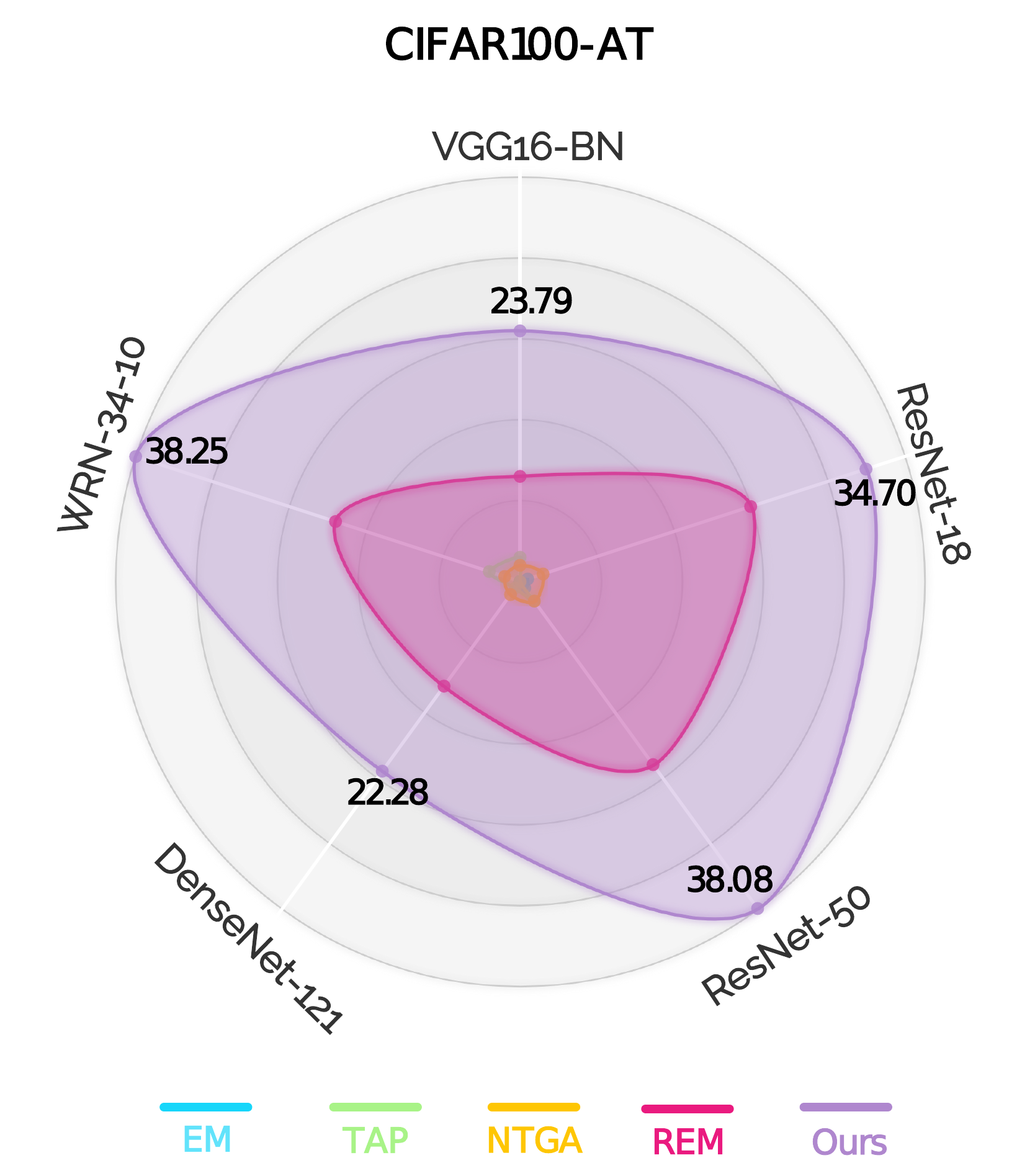}
\end{minipage}
\caption{Protective effects of five type unlearnable methods on CIFAR-10 and CIFAR-100. The protective perturbation radius of every method is set as $8/255$ and the adversarial training perturbation is set as $4/255$.}
\end{figure}

Additionally, we advocate that the surrogate model employed for generating unlearnable examples should exhibit robustness to ensure that the protective benefits conferred by unlearnable examples are not easily undermined by adversarial training. By synthesizing these insights, we ultimately present a comprehensive robust approach for generating generalized, robust unlearnable examples. As evidenced in Figure~\ref{fig:effects}, the generalization capability of our method across different datasets significantly surpasses that of existing methods.

In summary, our primary contributions can be outlined as follows: 
\begin{itemize}
    \item We highlight the shortcomings in the generalization capabilities of unlearnable examples present in prior approaches.
    \item For the first time, we introduce a method that focuses on the data collapse perspective to simultaneously enhance both the protective effects and generalization abilities of unlearnable examples.
    \item We introduce a method to match the gradient of the data distribution, which enables us to push the images to collapse together. 
    \item Extensive experiments have verified our method outperforms other unlearnable methods both in protection effects and generalization ability.
\end{itemize}

\section{Related Work}
\subsection{Poisoning Attacks}
Data poisoning attacks are strategically designed to impair the training process of a model by introducing noise into the training data, consequently leading to significant testing errors on specific or previously unseen samples during the evaluation phase. Backdoor attacks, a widespread form of poisoning attack, commonly involve incorporating triggers into training samples, which in turn cause the model to misclassify images with these triggers during testing~\cite{backdoor2021-1, backdoor2020-2, backdoor2020-3}. Notably, the effects of such attacks are generally confined to samples containing embedded trigger patterns, ensuring that clean samples remain unaffected and are accurately classified~\cite{backdoor2017-5, backdoor2018-4}.

\subsection{Unlearnable Methods}
Unlearnable methods aim to protect data from unauthorized access by introducing imperceptible unlearnable noise. Data samples affected by these techniques are termed as unlearnable examples. When deep neural networks are exposed to training with unlearnable examples, their performance on conventional test samples deteriorates, resembling random guessing outcomes. Consequently, this ensures an enhanced level of data protection~\cite{EM2021}.

\textbf{Non-robust Methods.}
\label{sec:non_robust_methods}
This type of method usually generates non-robust unlearnable examples, which means that these non-robust unlearnable examples can easily be corrupted by adversarial training and lose their protective effects, such as LSP~\cite{LSP2022}, CUDA~\cite{CUDA2023}, TAP~\cite{TAP2021}, NTGA~\cite{NTGA2021}, and EM~\cite{EM2021}. 

LSP~\cite{LSP2022} and CUDA~\cite{CUDA2023} generate unlearnable noise directly at the pixel level, as opposed to the feature level. As a result, methods within this category do not necessitate any feature information on clean data, leading to no significant relationship with data features. Due to their pixel-level approach, these methods exhibit remarkable efficiency. Nonetheless, the inherent design principle underlying this approach makes unlearnable examples susceptible to feature-based defense techniques, such as adversarial training, presenting a fundamental limitation that cannot be easily mitigated. 

TAP~\cite{TAP2021}, NTGA~\cite{NTGA2021}, and EM~\cite{EM2021} entail training surrogate models as non-robust models that learn non-robust features. As a result, the unlearnable noise generated through this strategy exclusively targets poisoned models subjected to standard training, merely preventing these models from learning standard data features. However, when the poisoned models undergo adversarial training, the protective effects conferred by these methods become compromised.


\textbf{Robust Methods.}
This type of method means the trained surrogate models are robust models that learned robust features. The only work is REM~\cite{REM2022}. REM~\cite{REM2022} regards that the poisoned model undergoing adversarial training learns knowledge from the adversarial examples. Therefore, REM~\cite{REM2022} generates unlearnable noise for the adversarial examples instead of the clean data, making unlearnable noise remains protective effects against adversarial training. 

Consider a dataset $\mathcal{D} = \{(x_1, y_1), (x_2, y_2), \dots, (x_n, y_n)\}$ comprising $n$ samples, where $x_i \in \mathcal{X}$ represents the $i$-th sample and $y_i \in \mathcal{Y} = \{1, \dots, K\}$ denotes the corresponding label. A parameterized machine learning model can be expressed as $f_\theta: \mathcal{X} \rightarrow \mathcal{Y}$, where $\theta \in \Omega$ signifies the model parameter. Additionally, let $\ell$ represent a loss function.
Then, the training objective of the robust noise generator~\cite{REM2022} is as follows:
\begin{equation}
\label{equ:rem}
    \min_{\theta} \frac{1}{n} \sum_{i=1}^n \min_{\|\delta^u_i\|\leq \rho_u} \max_{\|\delta^a_i\| \leq \rho_a} \ell(f'_\theta(x_i+\delta^u_i+\delta^a_i),y_i)
\end{equation} 
where $\rho_u$ is the defensive perturbation radius that forces the generated robust unlearnable noise to be imperceptible. 

Despite the partial protection REM~\cite{REM2022} offered against adversarial training, its generalization capability remains limited due to the excessive reliance on the surrogate models when generating unlearnable examples. 

\section{Methodology}
\subsection{Motivation}
\label{sec:motivation}
Current unlearnable methods have yet to conduct a comprehensive examination of the properties of unlearnable examples in impeding models from learning specific information. As discussed in section~\ref{sec:non_robust_methods}, unlearnable examples generated by non-robust methods are susceptible to adversarial training efforts. While REM~\cite{REM2022} stands as the sole robust method, it scarcely considers the generalizability of unlearnable examples. In light of this, our objective is to propose a generalizable and robust method for generating generalizable and robust unlearnable examples.
Since the information concealed in data is fundamentally dependent upon the data itself, we strive to induce data collapse. To achieve this, we aim to consolidate the same category of data, thereby reducing the information hidden within the data distribution. Our approach represents the first attempt to generate unlearnable examples from a data collapse perspective, leading to a more comprehensive and robust solution.

Additionally, upon examining Eq \ref{equ:rem}, it becomes evident that the trained surrogate model of REM follows a standard training procedure. As a result, the surrogate model primarily acquires non-robust features rather than robust features~\cite{ST2019}. 

\begin{figure}[htbp]
\hspace{-6mm}
\begin{minipage}[t]{0.45\linewidth}
\centering
\includegraphics[scale=0.15]{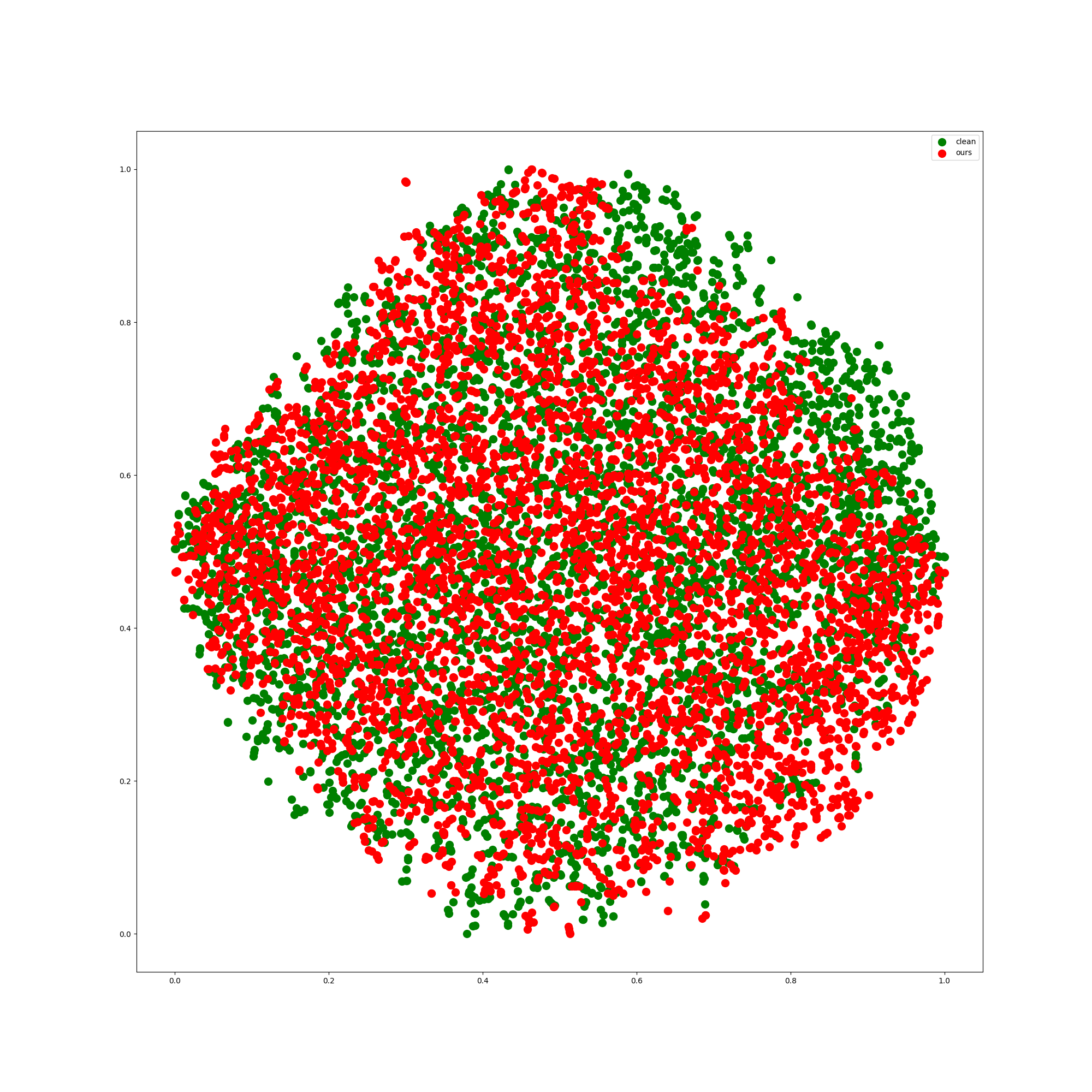}
\end{minipage}%
\hspace{5mm}
\begin{minipage}[t]{0.45\linewidth}
\centering
\includegraphics[scale=0.15]{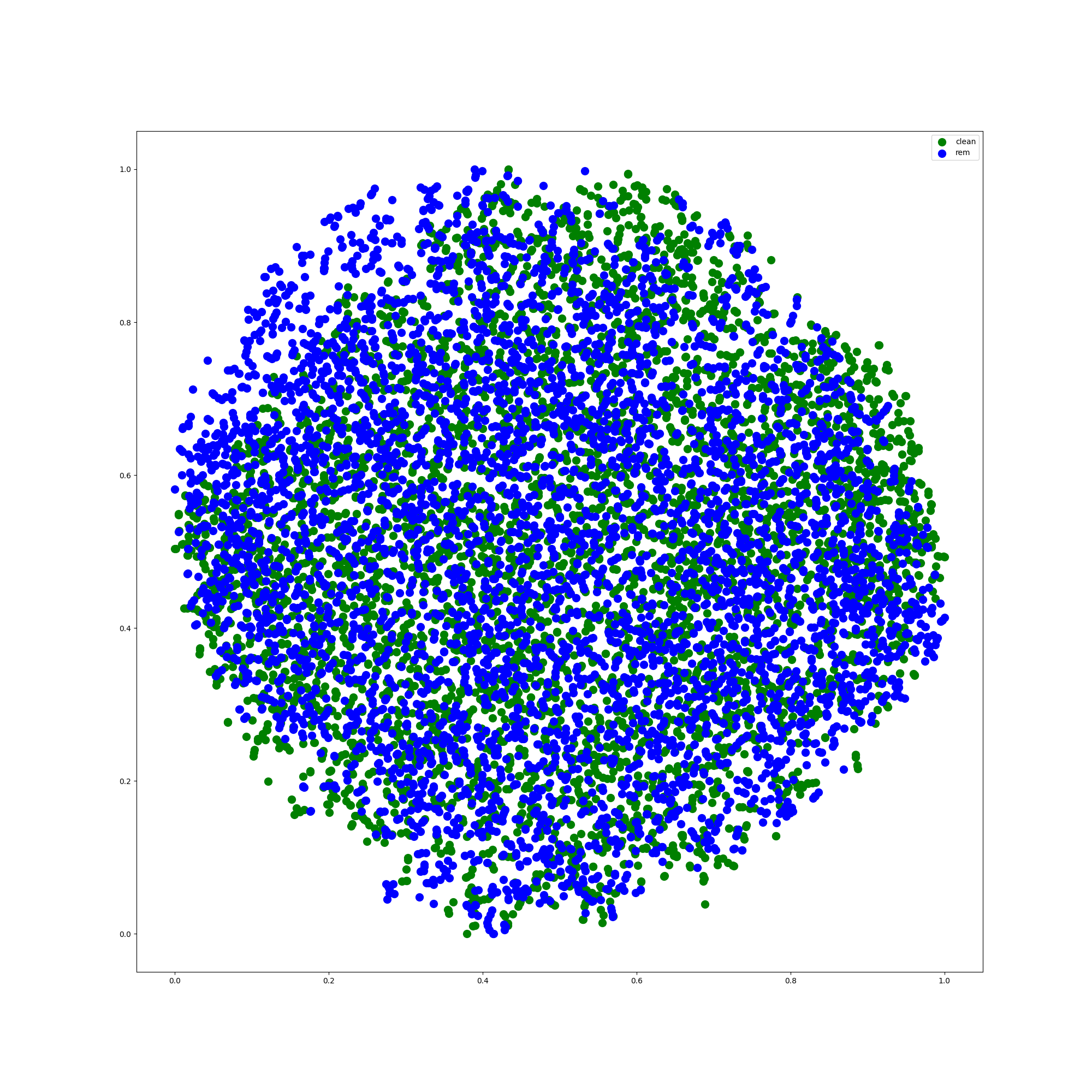}
\end{minipage}
\caption{t-SNE visualizations. We illustrate the t-SNE plots for clean (green)
examples, unlearnable examples generated by our method (red), and unlearnable examples generated by REM (blue). }
\end{figure}

\subsection{Direction of Data Distribution}

According to Stochastic Gradient Langevin Dynamics (SGLD)~\cite{LZW1, LZW2}, the iterative updating rule is given by:
\begin{equation}
    x_t = x_{t-1} + \alpha \cdot \nabla_{x_{t-1}}\log p_D(x_{t-1}|y) + \sqrt{2\alpha}\cdot \epsilon 
\end{equation}
As $\alpha \rightarrow 0$ and $t \rightarrow \infty$, $x_t$ converges to a sample from $p_D(x|y)$. Updating the SGLD process along the direction of $\nabla_{x\log p_D(x|y)}$ transitions the image towards the same image without random noise. 
Inspired by this function, once we get the gradient of the ground truth data distribution, we can easily manipulate data distribution. Updating the SGLD process along the opposite direction of $\nabla_x p_D(x|y)$ can move the image away from the distribution $p_D(x|y)$ while updating the SGLD process along the direction of $\nabla_x p_D(x|y)$ can move the image gather together. 

In this section, we propose to match the gradient of the log ground truth class-conditional data distribution $s_{\theta}(x,y) = \nabla_x \log p_{\theta}(x|y)$. Specifically, in our setting, if we estimate the gradient of the ground truth data distribution well, the process of SGLD sampling can move the samples of the same category more concentrated, which can hide information from the perspective of data distribution. 

We need to train an estimator of the gradient of the log ground truth class conditional data distribution. The optimization object is as follows:
\begin{equation}
\label{equ:ori_score}
    \frac{1}{2} \mathbb{E}_{p_D} [\norm{s_\theta(x, y) - \nabla_x \log p_D(x|y)}_2^2]
\end{equation}

Since we don't know the real data distribution, we need some processing to make this real data distribution into a distribution that we know explicitly. The approach is to add Gaussian noise to the original data so that the result satisfies our pre-defined distribution. In this way, we know the probability density now, and can train using a practice like that of equation \ref{equ:ori_score}. Denote training sample is $x$, the noise added sample is $\tilde{x}$, the predefined  distribution is $q_\sigma(\tilde{x}|x,y) \sim N\left(\tilde{x};x,y,\sigma^2 I\right)$. Then, the distribution of the data after noise addition is expressed as $q_\sigma(\tilde{x}) \equiv \int q_\sigma(\tilde{x}|x,y) p_D(x,y)dx$. Then, the optimization object is changed as follows: 
\begin{equation}
\frac{1}{2} \mathbb{E}_{q_\sigma(\tilde{x} \mid x,y) p_D(x)}\left[\left\|s_\theta(\tilde{x},y)-\nabla_{\tilde{x}} \log q_{\sigma}(\tilde{x}|x,y)\right\|_2^2\right]
\end{equation}
The derivation of $\nabla_{\tilde{x}} \log q_{\sigma}(\tilde{x}|x,y)$ is $-\frac{\tilde{x}-x}{\sigma^2}$. The final optimization object is:
\begin{equation}
\frac{1}{2} \mathbb{E}_{q_\sigma(\tilde{x} \mid x,y) p_D(x)}\left[\left\|s_\theta(\tilde{x},y)+\frac{\tilde{x}-x}{\sigma^2}\right\|_2^2\right]
\end{equation}

In this approach, we sample random noise, $\epsilon$, from the standard Gaussian distribution $N(0, I)$, and then multiply it by our pre-defined $\sigma$ before adding it to the sample $x$. Consequently, this yields the noisy sample $\tilde{x}=x+\sigma \epsilon$. According to the Gaussian distribution's properties, the noise-added conditional distribution $q_\sigma(\tilde{x}|x,y)$ will satisfy $N\left(\tilde{x};x,y,\sigma^2 I\right)$. 


In this way, we train an independent log gradient estimator of the gradient of the log ground truth class-conditional data distribution $\nabla_x \log p_{\theta}(x|y)$.

\subsection{Generalizable Robust Noise Generator}
Utilizing the log gradient estimator, we can alter the data distribution to effectively conceal the inherent information. By guiding the model to learn less information instead of misinformation, we direct the data towards the gradient, causing it to eventually collapse. In the ideal scenario, similar data will collapse at the same location, reducing variability between analogous data. 

Upon scrutiny of the training objectives for existing robust availability attacks, REM~\cite{REM2022}, it is apparent that the training procedure for surrogate models follows a standard protocol. Consequently, the surrogate model primarily captures non-robust features rather than robust features, as illustrated by the study conducted by~\cite{ST2019}. Although prior analyses suggest that such unlearnable examples may not provide sufficient protection against adversarial training, the effectiveness of REM~\cite{REM2022} warrants further exploration. A potential explanation for the success of REM could be the implementation of an adversarial training procedure, which is incorporated into their code but not explicitly mentioned in their theoretical framework. We posit that the robustness arises from the surrogate model's ability to learn the data's robust features. Therefore, by integrating these insights, we endeavor to train a generalizable and robust surrogate model. Our robust unlearnable examples' generator training encompasses two stages:



\begin{enumerate}
    \item[a)] Generate generalized unlearnable examples $x'$ from the surrogate model $f'_{\theta}$ by:
    \begin{equation}
        \delta_i^u =\min_{||\delta_i^u||\leq \rho_u} ||s_{\theta}(x_i+\delta_i^u, y_i)|| + \min_{||\delta_i^u||\leq \rho_u} \ell(f'_{\theta}(x_i+\delta_i^u),y_i) 
        \label{stage-one}
    \end{equation}
    \item[b)] Perform adversarial training on the surrogate model $f'_{\theta}$ to extract robust features from unlearnable examples:
    \begin{equation}
        \min_{\theta} \frac{1}{n}\sum_{i=1}^{n} \max_{||\delta_i^a|| \leq \rho_a}\ell(f'_{\theta}(x_i+\delta_i^u+\delta_i^a),y_i)
        \label{stage-two}
    \end{equation}
\end{enumerate}

Minimizing the loss $\ell(f'_{\theta}(x_i+\delta_i^u),y_i)$, making the loss too small to update the parameters, while minimizing $||s_{\theta}(x_i+\delta_i^u, y_i)||$ alters the distribution, resulting in data collapse. Further details are provided in Algorithm~\ref{algo:m2_noise_gen}. A generalized robust unlearnable example $(x', y)$ is crafted by adding the noise generated by the trained generalized robust noise generator $f'_{\theta}$ to its clean counterpart $(x, y)$. The resulting unlearnable example is denoted as $x' = x + \delta^u$.
The details are represented in Algorithm~\ref{algo:m2_noise_gen}.

\begin{algorithm}[th]
\caption{Training our unlearnable examples' generator}
\label{algo:m2_noise_gen}
\begin{algorithmic}[1]
\Require

Training data set $\mathcal{D}$,
training iteration $M$, 
score estimator $s_{\theta}$,
learning rate $\eta$\\
PGD parameters $\rho_u$, $\alpha_u$, $\alpha_s$ and $K_u$ for step one, \\
PGD parameters $\rho_a$, $\alpha_a$ and $K_a$ for step two. 
\Ensure unlearnable examples generator $f'_\theta$.
    \State Initialize source model parameter $\theta$.

   
    \For{$i$ \textbf{in} $1, \cdots, M$} 
        \State Sample a minibatch $(x, y) \sim \mathcal{D}$.
        \State Initialize $\delta^u$. 
        \For{$k$ \textbf{in} $1,\cdots, K_u$}   \tikzmark{a}
            \State $g_k \leftarrow  \frac{\partial}{\partial \delta^u} \ell(f'_\theta(x+\delta^u), y)$ \Comment{data collapse}
            \State $g_k \leftarrow  \frac{\partial}{\partial \delta^u} ||(s_{\theta}(x+\delta^u, y)||$ \Comment{minimize error}
            \State $\delta^u \leftarrow \prod_{\|\delta\|\leq\rho_u} \left( \delta^u - \alpha_u \cdot \mathrm{sign}(g_k) -\alpha_s \cdot \mathrm{sign}(s_k) \right)$ 
        \EndFor  \tikzmark{b} 
        \For{$k$ \textbf{in} $1,\cdots, K_a$}
            \State $g_k \leftarrow  \frac{\partial}{\partial \delta^u} \ell(f'_\theta(x+\delta^u+\delta^a), y)$
            \State $\delta^a \leftarrow \prod_{\|\delta\|\leq\rho_a} \left( \delta^a - \alpha_a \cdot \mathrm{sign}(g_k) \right)$
        \EndFor
        \State $g_k \leftarrow \frac{\partial}{\partial \theta} \ell(f'_\theta(x+\delta^u+\delta^a), y) $
        \State $\theta \leftarrow \theta - \eta \cdot g_k$
    
    \EndFor        
    \State \Return $f'_\theta$
    
\end{algorithmic}
\end{algorithm}

\section{Experiments}
In this section, we have conducted sufficient experiments to demonstrate the generalizability and effectiveness of unlearnable examples generated by our method from various aspects which can prevent models from learning knowledge through adversarial training.

\subsection{Experiment Setup}
\textbf{Datasets.} 
To verify the effectiveness of our method on images of different categories and sizes, three commonly used datasets, namely CIFAR-10, CIFAR-100~\cite{CIFAR10}, and ImageNet subset~\cite{ImageNet2015} (consists of the first 100 classes), are used in our experiments. The data augmentation technique~\cite{DataAugment2019} is adopted in every experiment.

\textbf{Surrogate Models.}
Following~\cite{EM2021} and~\cite{REM2022}, we employ ResNet-18\cite{Resnet18} as the surrogate model $f'_{\theta}$ for trainging our unlearnable examples generator with Eq.~(\ref{stage-one}) and Eq.~(\ref{stage-two}). The $L_\infty$-bounded perturbation  $\|\delta_u\|_\infty \leq \rho_u$ is adopted in our experiments. Additionally, we also use other surrogate models, including VGG-16\cite{VGG11}, ResNet-50\cite{Resnet18}, and DenseNet-121\cite{Densenet121}, to test the generalizability of our method.


\textbf{Noise Test.}
Noise generated by our method is tested both on standard training and adversarial training~\cite{AdversarialTraining2018}. We focus on $L_\infty$-bounded noise $\|\rho_a\|_\infty \leq \rho_a$ in adversarial training. We conduct adversarial training on unlearnable examples created by our method with different models, including VGG-16~\cite{VGG11}, ResNet-18, ResNet-50~\cite{Resnet18}, DenseNet-121~\cite{Densenet121}, and wide ResNet-34-10~\cite{WRN2016}. 
Note that when $\rho_a$ takes $0$, the adversarial training degenerates to the standard training.

\textbf{Metric.}
We evaluate the data protection ability of defensive noise by measuring the test accuracy of the model. A low test accuracy indicates that the model has learned little from the unlearnable examples, implying a strong protection ability from the noise.

\subsection{Efftiveness of Our Method}
\subsubsection{Different adversarial training perturbation radius.}
To assess the generalizability and robustness against adversarial training, we first introduce unlearnable noise to the entire training set, generating unlearnable examples. 
We train models using different adversarial training perturbation radii $\rho_a$ on these unlearnable examples. The adversarial training perturbations $\rho_a$ range from $0/255$ to $4/255$. It is important to note that when $\rho_a=0$, the models are trained using the standard training method.
The unlearnable noise perturbation radius, denoted as $\rho_u$, is set to $8/255$ for all noise-generating methods and the adversarial perturbation radius $\rho_a$ is set as $4/255$ for REM~\cite{REM2022} and our method. 
Table~\ref{tab:diff_radius} reports the accuracies of the trained models on the unlearnable examples. 

\begin{table}[h]
\centering
\caption{Test accuracy (\%) of models trained on unlearnable examples generated by different defensive noises via adversarial training with different perturbation radii.}
\label{tab:diff_radius}
\scriptsize
\begin{tabular}{c c c c c c c c}
\toprule
\multirow{2}{1.2cm}{\centering Dataset} & \multirow{3}{1.2cm}{\centering Adv. Train. \\ $\rho_a$} & \multirow{2}{0.8cm}{\centering Clean} & \multirow{2}{0.8cm}{\centering EM} & \multirow{2}{0.8cm}{\centering TAP} & \multirow{2}{0.8cm}{\centering NTGA} &  {\centering REM} & {\centering Ours} \\
\cmidrule(lr){7-8}
& & & & & & $\rho_a = 4/255$ & $\rho_a = 4/255$ \\
\midrule
\multirow{5}{1.2cm}{\centering CIFAR-10}
& $0$     & 94.66 & 13.20 & 22.51 & 16.27 & 22.93 & \textbf{10.47} \\
& $1/255$ & 93.74 & 22.08 & 92.16 & 41.53 & 30.00 & \textbf{14.33}  \\
& $2/255$ & 92.37 & 71.43 & 90.53 & 85.13 & 30.04 & \textbf{18.80} \\
& $3/255$ & 90.90 & 87.71 & 89.55 & 89.41 & 31.75 & \textbf{24.44} \\
& $4/255$ & 89.51 & 88.62 & 88.02 & 88.96 & 48.16 & \textbf{29.20} \\
\midrule
\multirow{5}{1.2cm}{\centering CIFAR-100}
& $0$     & 76.27 &  \textbf{1.60} & 13.75 & 3.22 & 11.63 & 2.70  \\
& $1/255$ & 71.90 & 71.47 & 70.03 & 65.74 & 14.48 & \textbf{4.07}  \\ 
& $2/255$ & 68.91 & 68.49 & 66.91 & 66.53 & 16.60 & \textbf{5.21}  \\ 
& $3/255$ & 66.45 & 65.66 & 64.30 & 64.80 & 20.70 & \textbf{12.06}  \\ 
& $4/255$ & 64.50 & 63.43 & 62.39 & 62.44 & 27.35 & \textbf{22.35} \\ 
\midrule
\multirow{5}{1.2cm}{\centering ImageNet Subset}
& $0$     & 80.66 & \textbf{1.26} &  9.10 & 8.42 & 13.74 & 7.96 \\
& $1/255$ & 76.20 & 74.88 & 75.14 & 63.28 & 21.58 & \textbf{13.20} \\
& $2/255$ & 72.52 & 71.74 & 70.56 & 66.96 & 29.40 & \textbf{17.00} \\
& $3/255$ & 69.68 & 66.90 & 67.64 & 65.98 & 35.76 & \textbf{21.42} \\
& $4/255$ & 66.62 & 63.40 & 63.56 & 63.06 & 41.66 & \textbf{29.28} \\
\bottomrule
\end{tabular}
\vspace{-4mm}
\end{table}

The surrogate models in Table~\ref{tab:diff_radius} are ResNet-18~\cite{Resnet18}. 
For adversarial training, we can find that even a very small adversarial training perturbation radius of $2/255$ can damage the protecting effects of TAP~\cite{TAP2021}, NTGA~\cite{NTGA2021}, and EM\cite{EM2021}. 
But our method can be generalized to different perturbations, always maintaining a good protection effect. Even in the large perturbation of $4/255$, our method can outperform the SOTA method over $5\% \sim 19\%$.   
For standard training, our method also performer well than other methods. 
The three datasets are of different sizes and classes. 
Even in this complex situation, our method can always keep significant protective effects over other methods, especially in adversarial training.
Totally, these experiments demonstrate that our method can be applied to different datasets and different adversarial training perturbation radii while keeping a significant protective effect.

\subsubsection{Generalizability}

\textbf{Different model architectures.}
Until now, we have only conducted adversarial training with ResNet-18, which is as same as the source model in the defensive noise generation.
We now evaluate the generalizability of the unlearnable examples generated by our method under different adversarial training models. 
Specifically, we conduct adversarial training with a perturbation radius of $4/255$ and five different types of models, including VGG-16, ResNet-18, ResNet-50, DenseNet-121, and wide ResNet-34-10, on data that is protected by noise generated via ResNet-18. The defensive perturbation radius $\rho_u$ of every type of defensive noise is set as $8/255$. 
Table~\ref{tab:diff_arch_at} presents the test accuracies of the trained models on CIFAR-10 and CIFAR-100.

\begin{table}[h]
\centering
\caption{Test accuracy (\%) of different types of models on CIFAR-10 and CIFAR-100 datasets.}
\label{tab:diff_arch_at}
\scriptsize
\begin{tabular}{c c c c c c}
\toprule
\multirow{2}{1.2cm}{\centering Dataset} & \multirow{2}{0.8cm}{\centering Model} & \multirow{2}{0.8cm}{\centering Clean} & \multirow{2}{0.8cm}{\centering EM} & \multicolumn{1}{c}{\centering REM} & \multicolumn{1}{c}{\centering Ours} \\
\cmidrule(lr){5-6}
& & & & $\rho_a = 4/255$ & $4/255$ \\
\midrule
\multirow{5}{1.2cm}{\centering CIFAR-10}
& VGG-16    & 87.51 & 87.24 & 65.23 & \textbf{37.55}    \\
& RN-18     & 89.51 & 88.62 & 48.16 & \textbf{29.20}    \\
& RN-50     & 89.79 & 89.66 & 40.65 & \textbf{25.63}    \\
& DN-121    & 83.27 & 81.77 & 82.38 & \textbf{64.46}    \\
& WRN-34-10 & 91.21 & 79.87 & 48.39 & \textbf{22.98}    \\

\midrule
\multirow{5}{1.2cm}{\centering CIFAR-100}
& VGG-16    & 57.46 & 56.94 & 58.07 & \textbf{55.22}    \\
& RN-18     & 63.43 & 63.43 & 27.35 & \textbf{22.35}    \\
& RN-50     & 66.93 & 66.43 & 26.03 & \textbf{21.60}    \\
& DN-121    & 53.73 & 53.52 & 56.63 & \textbf{52.32}    \\
& WRN-34-10 & 68.64 & 68.27 & 27.71 & \textbf{19.44}    \\
\bottomrule
\end{tabular}
\vspace{-3mm}
\end{table}

Table~\ref{tab:diff_arch_at} shows that our unlearnable examples generated from ResNet-18 can effectively protect data against various adversarially trained models and outperforms other methods by a significant margin. 


\textbf{Different Noise generator} 

Thus far, all our unlearnable examples have been generated using the ResNet-18 surrogate model. It is important to note that, as detailed in section~\ref{sec:motivation}, the enhancement of unlearnable examples' generalizability is achieved by data collapse, meaning it is not dependent on the choice of surrogate models. In previous experiments, the generalizability of unlearnable examples generated by ResNet-18 was verified. To further substantiate the generalizability of our method, we seek to assess the effectiveness of unlearnable examples generated by other surrogate models.

We examine four surrogate models, including VGG-16, ResNet-18, ResNet-50, and DenseNet-121. Each type of unlearnable example is tested on five models, comprising VGG-16, ResNet-18, ResNet-50, DenseNet-121, and WRN-34-10. Experiments are conducted under both standard and adversarial training scenarios. The adversarial training perturbation radius is set to $4/255$, while the defensive perturbation radius $\rho_u$ for each unlearnable method is set to $8/255$.

Test accuracies of standard training models on CIFAR-10 and CIFAR-100 are presented in Table~\ref{tab:trans_cifar10_st} and Table~\ref{tab:trans_cifar100_st}. Likewise, test accuracies of adversarial training models on CIFAR-10 and CIFAR-100 can be found in Table~\ref{tab:trans_cifar10_at} and Table~\ref{tab:trans_cifar100_at}.

\begin{table}[h]
\centering
\caption{Test accuracy (\%) of models standardly trained on unlearnable examples generated by different surrogate models on CIFAR-10.
The defensive perturbation radius $\rho_u$ of every type of defensive noise is set as $8/255$.}
\label{tab:trans_cifar10_st}
\scriptsize
\begin{tabular}{c c c c c c c c }
\toprule

{\centering Surrogate Model} & {\centering Method} & {\centering VGG-16} & {\centering ResNet-18} & {\centering ResNet-50} & {\centering DenseNet-121} & {\centering WRN-34-10} & {\centering Average} \\
\midrule
\multirow{3}{1.2cm}{\centering VGG-16}
& EM        & 70.35 & 77.84 & 62.63 & 71.24 & 66.43 & 69.70    \\
& REM       & 46.27 & 45.25 & 40.29 & 43.29 & 42.95 & 43.61    \\
& Ours        & \textbf{26.59} & \textbf{26.23} & \textbf{29.41} & \textbf{32.98} & \textbf{30.25} & \textbf{29.10}    \\
\midrule
\multirow{3}{1.2cm}{\centering ResNet-18}
& EM        & 15.70 & 13.20 & 10.19 & 14.59 & 10.56 & 12.85   \\
& REM       & 23.55 & 22.93 & 22.98 & 28.47 & 23.19 & 24.22    \\
& Ours        & \textbf{10.97} & \textbf{10.47} & \textbf{10.04} & \textbf{12.26} & \textbf{10.75} & \textbf{10.90}    \\
\midrule
\multirow{3}{1.2cm}{\centering ResNet-50}
& EM        & 18.77 & 17.89 & 19.31 & 26.08 & 19.41 & 20.29    \\
& REM       & 24.64 & 23.22 & 23.91 & 30.04 & 22.82 & 24.93    \\
& Ours        & \textbf{10.01} & \textbf{11.23} & \textbf{11.51} & \textbf{11.91} & \textbf{10.41} & \textbf{11.01}\\
\midrule
\multirow{3}{1.2cm}{\centering DenseNet-121}
& EM     & 17.67 & 12.30 & 11.76 & 18.85 & 11.23 & 14.36       \\
& REM    & 35.46 & 35.91 & 33.09 & 33.26 & 33.49 & 34.24       \\
& Ours     & \textbf{16.14} & \textbf{11.30} & \textbf{11.10} & \textbf{18.38} & \textbf{10.91} & \textbf{13.57}  \\
\bottomrule
\end{tabular}
\vspace{-3mm}
\end{table}

\begin{table}[h]
\centering
\caption{Test accuracy (\%) of models standardly trained on unlearnable examples generated by different surrogate models on CIFAR-100.
The defensive perturbation radius $\rho_u$ of every type of defensive noise is set as $8/255$.}
\label{tab:trans_cifar100_st}
\scriptsize
\begin{tabular}{c c c c c c c c }
\toprule
{\centering Surrogate Model} & {\centering Method} & {\centering VGG-16} & {\centering ResNet-18} & {\centering ResNet-50} & {\centering DenseNet-121} & {\centering WRN-34-10} & {\centering Average} \\
\midrule
\multirow{3}{1.2cm}{\centering VGG-16}
& EM        & 17.45 & 21.79 & 28.94	& 61.85	& 34.50 & 32.91 \\
& REM       & 11.89	& 21.08	& 19.76	& 17.95	& 26.72 & 19.48 \\
& Ours        & \textbf{9.27} & \textbf{11.50} & \textbf{11.00} & \textbf{13.86} & \textbf{10.97} & \textbf{11.32}\\
\midrule
\multirow{3}{1.2cm}{\centering ResNet-18}
& EM        & 4.59 & \textbf{1.60} & 4.77 & 6.22 & 3.59 & 4.15 \\
& REM       & 9.15 & 11.63 & 9.23 & 13.06 & 11.74 & 10.97 \\
& Ours        & \textbf{4.38} & 2.70 & \textbf{3.30} & \textbf{7.84} & \textbf{2.50} & \textbf{4.14} \\
\midrule
\multirow{3}{1.2cm}{\centering ResNet-50}
& EM        & 69.10 & 74.79 & 74.16 & 65.18 & 76.71 & 71.99 \\
& REM       & 10.24 & 10.53 & 9.43 & 23.28 & 10.15 & 12.73 \\
& Ours        & \textbf{4.21} & \textbf{6.92} & \textbf{6.86} & \textbf{6.38} & \textbf{7.63} & \textbf{6.40} \\
\midrule
\multirow{3}{1.2cm}{\centering DenseNet-121}
& EM        & \textbf{1.00} & 7.84 & 7.53 & 64.76 & 11.15 & 18.46 \\
& REM       & 16.62 & 20.28 & 18.88 & 17.96 & 20.91 & 18.93 \\
& Ours        & 3.52 & \textbf{5.07} & \textbf{5.20} & \textbf{8.32} & \textbf{10.22} & \textbf{6.47} \\
\bottomrule
\end{tabular}
\vspace{-3mm}
\end{table}

As illustrated in Table~\ref{tab:trans_cifar10_st} and Table~\ref{tab:trans_cifar100_st}, it is evident that our method can be applied to various surrogate models while consistently achieving superior protective effects compared to other approaches. Notably, other methods exhibit sensitivity to the choice of surrogate models; however, our method remains stable and effective regardless of the surrogate model's learning ability.

\begin{table}[h]
\centering
\caption{Test accuracy (\%) of models adversarially trained on unlearnable examples generated by different surrogate models on CIFAR-10.
The adversarial training perturbation radius is set as $4/255$. The defensive perturbation radius $\rho_u$ of every type of defensive noise is set as $8/255$.}
\label{tab:trans_cifar10_at}
\scriptsize
\begin{tabular}{c c c c c c c c }
\toprule

{\centering Surrogate Model} & {\centering Method} & {\centering VGG-16} & {\centering ResNet-18} & {\centering ResNet-50} & {\centering DenseNet-121} & {\centering WRN-34-10} & {\centering Average} \\
\midrule
\multirow{3}{1.2cm}{\centering VGG-16}
& EM        & 87.75 & 89.21 & 90.19 & 83.58 & 90.83 & 88.31    \\
& REM       & 73.60 & 74.73 & 74.16 & 77.63 & 74.94 & 75.01    \\
& Ours        & \textbf{64.72} & \textbf{54.17} & \textbf{62.35} & \textbf{70.21} & \textbf{62.36} & \textbf{62.76}    \\
\midrule
\multirow{3}{1.2cm}{\centering ResNet-18}
& EM        & 87.24 & 88.62 & 89.66 & 81.77 & 79.87 & 85.43    \\
& REM       & 65.23 & 48.16 & 40.65 & 82.38 & 48.39 & 56.96    \\
& Ours        & \textbf{37.55} & \textbf{29.20} & \textbf{25.63} & \textbf{64.46} & \textbf{22.98} & \textbf{35.96}    \\
\midrule
\multirow{3}{1.2cm}{\centering ResNet-50}
& EM        & 87.57	& 89.17	& 89.83	& 82.64 & 90.68 & 87.98    \\
& REM       & 51.88 & 44.27 & 37.79 & 82.01 & 42.09 & 51.69    \\
& Ours        & \textbf{42.64} & \textbf{30.04} & \textbf{28.78} & \textbf{74.42} & \textbf{30.44} & \textbf{41.26}\\
\midrule
\multirow{3}{1.2cm}{\centering DenseNet-121}
& EM     & 87.59 & 84.51 & 85.57 & 82.76 & 85.68 & 85.22       \\
& REM    & 67.30 & 69.62 & 66.42 & 60.51 & 72.09 & 67.19       \\
& Ours     & \textbf{35.88} & \textbf{36.61} & \textbf{31.39} & \textbf{39.35} & \textbf{34.78} & \textbf{35.60}  \\
\bottomrule
\end{tabular}
\vspace{-3mm}
\end{table}

\begin{table}[h]
\centering
\caption{Test accuracy (\%) of models adversarially trained on unlearnable examples generated by different surrogate models on CIFAR-100.
The adversarial training perturbation radius is set as $4/255$. The defensive perturbation radius $\rho_u$ of every type of defensive noise is set as $8/255$.}
\label{tab:trans_cifar100_at}
\scriptsize
\begin{tabular}{c c c c c c c c }
\toprule
{\centering Surrogate Model} & {\centering Method} & {\centering VGG-16} & {\centering ResNet-18} & {\centering ResNet-50} & {\centering DenseNet-121} & {\centering WRN-34-10} & {\centering Average} \\
\midrule
\multirow{3}{1.2cm}{\centering VGG-16}
& EM        & 57.33 & 63.55 & 65.44 & 53.45 & 68.23 & 61.60 \\
& REM       & 41.13 & 52.00 & 51.77 & 48.92 & 56.05 & 49.97 \\
& Ours        & \textbf{34.02}	& \textbf{46.66}	& \textbf{46.55}	& \textbf{41.93}	& \textbf{49.23} & \textbf{43.68}\\
\midrule
\multirow{3}{1.2cm}{\centering ResNet-18}
& EM        & 56.94 & 64.17 & 66.43 & 53.52 & 68.27 & 61.87 \\
& REM       & 58.07 & 27.35 & 26.03 & 56.63 & 27.71 & 39.16 \\
& Ours        & \textbf{55.22} & \textbf{22.35} & \textbf{21.60} & \textbf{52.32} & \textbf{19.44} & \textbf{34.27}\\
\midrule
\multirow{3}{1.2cm}{\centering ResNet-50}
& EM        & 56.82 & 64.19 & 66.93 & 54.51 & 68.56 & 62.20 \\
& REM       & 54.61 & 35.50 & 30.43 & 54.26 & 35.11 & 41.98 \\
& Ours        & \textbf{51.24} & \textbf{26.01} & \textbf{21.81} & \textbf{52.32} & \textbf{22.89} & \textbf{34.85} \\
\midrule
\multirow{3}{1.2cm}{\centering DenseNet-121}
& EM        & 57.39 & 63.73 & 66.37 & 54.62 & 68.43 & 62.11 \\
& REM       & 47.22 & 41.89 & 45.49 & 41.15 & 50.66 & 45.28 \\
& Ours        & \textbf{33.67} & \textbf{29.80} & \textbf{28.85} & \textbf{31.45} & \textbf{30.39} & \textbf{30.83} \\
\bottomrule
\end{tabular}
\vspace{-3mm}
\end{table}

As shown in Table~\ref{tab:trans_cifar10_at} and Table~\ref{tab:trans_cifar100_at}, we can find that our unlearnable examples are more robust than other methods despite the surrogate models. Moreover, the generalizability of unlearnable examples generated by our method is significant. The Average accuracies of our unlearnable examples can reach $5\% \sim 32\%$ down than other SOTA methods. 

Combining standard training and adversarial training on CIFAR-10 and CIFAR-100, our method achieves amazing protective effects. The good generalizability is attributed to the data collapse. 

\subsubsection{Different protection percentages.}
Realistically, there is a more challenging scenario, where only a part of the data is protected by the unlearnable noise, while the others are clean. 
Specifically, we randomly select a part of the training data from the whole training set, adding unlearnable noise to the selected data. We then conduct adversarial training with ResNet-18 on the mixed data and the remaining clean data. The detailed results are in supplementary materials.

\section{Conclusion}
In conclusion, we present, for the first time, a generalizable and robust unlearnable method based on the concept of data collapse. We introduce a common approach for estimating the gradient of the data distribution and render the data less informative through data collapse. Performed from the perspective of the data itself, data collapse is independent of any surrogate models and consequently enhances the generalization ability of unlearnable examples. Additionally, we highlight that utilizing a robust surrogate model can contribute to the robustness of unlearnable examples. By incorporating these insights, our method achieves noteworthy performance in generating generalizable, robust unlearnable examples.


\textbf{Limitations.} 
The proposed method in this paper requires the introduction of an adversarial training process to generate robust unlearnable samples. This leads to a significant computational cost when extending the method to large-scale datasets, such as ImageNet. As a result, the scalability of the proposed approach may be limited, especially when dealing with massive datasets.
Furthermore, the method necessitates the additional training of a score model to represent the data distribution. Compared to other approaches, this requirement increases the generation cost of protective noise to a certain extent. Future research could explore ways to optimize the process or develop alternative techniques that can achieve similar results with lower computational overhead.

{
\small
\bibliographystyle{unsrt}
\bibliography{refs}
}

\end{document}